\documentclass[smallextended]{svjour3}     % onecolumn (second format); required for SCW, see http://www.springer.com/economics/economic+theory/journal/355
\smartqed  % flush right qed marks, e.g. at end of proof
\usepackage[nice]{nicefrac}
\usepackage{graphicx}
\usepackage{wrapfig}
\newcommand\eat[1]{}
\usepackage{tikz}
\usepackage{booktabs}  %% nice tables
\usetikzlibrary{arrows}
 \journalname{Working Paper}
\usepackage{array,xspace,multirow,hhline,graphicx,xcolor,tikz,colortbl,tabularx,amsmath,amssymb,amsfonts}
\usepackage[round]{natbib}
\usetikzlibrary{shapes}
\usepackage{algorithm,algorithmic}
\usepackage{mathrsfs}
\usepackage{enumitem}
\setenumerate[1]{label=(\emph{\roman{*}}),ref=(\emph{\roman{*}}),leftmargin=*}

\usepackage{txfonts}
\usepackage{eucal} % consistency with minstable

\newcommand{\eg}{e.g.,\xspace}

\usepackage{tikz}

\newlength{\wordlength}

\newcommand{\Pref}[1][]{
	\ifthenelse{\equal{#1}{}}{\mathrel \succsim}{\mathop{\succsim_{#1}}}
}                                          
\newcommand{\sPref}[1][]{                  
	\ifthenelse{\equal{#1}{}}{\mathrel \succ}{\mathop{\succ_{#1}}}
}                                          
\newcommand{\Indiff}[1][]{                 
	\ifthenelse{\equal{#1}{}}{\mathrel \sim}{\mathop{\sim_{#1}}}
}
\newcommand{\prefset}[1][]{\ifthenelse{\equal{#1}{}}{\mathcal{\succsim}}{\mathcal{\succsim}_{#1}}}

\newcommand{\ml}[1][]{\ensuremath{\ifthenelse{\equal{#1}{}}{\mathit{ML}}{\mathit{ML}(#1)}}}
\newcommand{\sml}[1][]{\ensuremath{\ifthenelse{\equal{#1}{}}{\mathit{SML}}{\mathit{SML}(#1)}}}
\newcommand{\sd}[1][]{\ensuremath{\ifthenelse{\equal{#1}{}}{\mathit{SD}}{\mathit{SD}(#1)}}}
\newcommand{\rsd}[1][]{\ensuremath{\ifthenelse{\equal{#1}{}}{\mathit{RSD}}{\mathit{RSD}(#1)}}}
\newcommand{\rd}[1][]{\ensuremath{\ifthenelse{\equal{#1}{}}{\mathit{RD}}{\mathit{RD}(#1)}}}
\newcommand{\st}[1][]{\ensuremath{\ifthenelse{\equal{#1}{}}{\mathit{ST}}{\mathit{ST}(#1)}}}
\newcommand{\bd}[1][]{\ensuremath{\ifthenelse{\equal{#1}{}}{\mathit{BD}}{\mathit{BD}(#1)}}}
\newcommand{\pc}[1][]{\ensuremath{\ifthenelse{\equal{#1}{}}{\mathit{PC}}{\mathit{PC}(#1)}}}
\newcommand{\dl}[1][]{\ensuremath{\ifthenelse{\equal{#1}{}}{\mathit{DL}}{\mathit{DL}(#1)}}}
\newcommand{\ul}[1][]{\ensuremath{\ifthenelse{\equal{#1}{}}{\mathit{UL}}{\mathit{UL}(#1)}}}

\let\enumtemp=\enumerate
\def\enumerate{\enumtemp\itemsep 1pt}
\let\itemtemp=\itemize
\def\itemize{\itemtemp\itemsep 1pt}

\newcommand{\Omit}[1]{}

\begin{document}

% \title{A Probabilistic Approach to Voting, Assignment, and, Matching\thanks{}}

\title{A Probabilistic Approach to\\ Voting, Allocation, Matching, and Coalition Formation}%or Time-sharing 

	% \subtitle{Do you have a subtitle?\\ If so, write it here}


	\author{Haris Aziz}


	\institute{%
	  Haris Aziz \at
	  Data61, CSIRO and UNSW
	Sydney, Australia \\
	  Tel.: +61-2-9490\,59090 \\
	  Fax: +61-2-8306\,0405 \\
	  \email{haris.aziz@data61.csiro.au}	  
	         }

	\date{Received: date / Accepted: date}

\maketitle

\begin{abstract}
% Probabilistic social choice is a natural extension of the canonical voting setting where the outcome is a probability distribution over the alternative set. This entry gives an overview of probabilistic social choice functions in the literature and their relative merits according to the properties they satisfy.

Randomisation and time-sharing are some of the oldest methods to achieve fairness. I make a case that applying these approaches to social choice settings constitutes a powerful paradigm that deserves an extensive and thorough examination. %for several compelling reasons. 
I discuss challenges and opportunities in applying these approaches to settings including voting, allocation, matching, and coalition formation. 

\end{abstract}
	


\section{Introduction}

Suppose two agents have opposite preferences over the two possible social outcomes.
What should be a fair resolution for this problem? 

\smallskip
If the outcome is required to be deterministic, then it is patently unfair to one of the agents. However, one can regain fairness by at least three different approaches: (1) resort to randomisation so that each of the social outcomes has equal probability, (2) treat the outcomes as divisible and resort to time sharing where each social outcome has half of the time share or, (3) use a uniform frequency distribution if there will be multiple occurrences of the discrete outcomes. 
Mathematically, all three resolutions towards fairness are equivalent because the outcomes have equal probability, time-share, or frequency.
In the rest of the article, when I will describe a probabilistic approach to social choice, I will use it abstractly so as to model approaches (1), (2), and (3).

I argue that although a probabilistic approach has been applied in several social choice settings in both theory and practice~\citep[see e.g., ][]{Ston11a}, there is potential to revisit fundamental social choice settings such as voting, allocation, matching, and coalition formation with this powerful paradigm. Considering the natural aversion of many people to important decisions being based on the toss of the coin, such an approach may be especially useful for time and budget sharing scenarios.
A related chapter in this book is by \citet{Bran17b}.

\section{A case for probabilistic social choice}

I first list some of the compelling reasons why probabilistic social choice is a powerful and useful approach. 

\begin{enumerate}
	\item \textbf{Modeling time-sharing} Some of the settings and their corresponding results in the literature ignore the possibility of time-sharing. For example, several results in voting suppose that only one alternative is selected. However, the voting could be about deciding the fraction of time different genres of music are played on radio. Similarly most of the results in matching and coalition formation suppose that agents form exclusively one coalition or set of partnerships~\citep{Manl13a}. Many of these results need to be re-examined when we allow the flexibility of time-sharing.%~\citep{AtSe11a}. 
	\item \textbf{Participatory Budgeting} Voting can also be used to decide on which projects get how much budget~(see e.g.,~\citep{FGM16a,ABM17a}). The approach is getting traction as grassroots participatory budgeting movements grow in stature~\citep{Caba04a}. In this context, a probabilistic social choice view is useful because the probability of an alternative can represent the fraction of the budget allocated to it. 
	\item \textbf{Achieving fairness} As explained in the example above, a probabilistic or time-sharing approach to social choice is geared towards achieving fairness. 
	In the example above, no deterministic mechanism can  simultaneously be anonymous (treat agents symmetrically) and neutral (treat social outcomes symmetrically). On the other hand, a probabilistic approach easily overcomes this impossibility. % Randomization allows agents to be treated in a symmetric manner.
Suppose that in the example, the two social outcomes are allocating one item each to the agents where one item is valuable to both and the second item is useless to both. Then the only way to avoid envy is to use a probabilistic approach in a broad sense.

	\item \textbf{Incentivizing participation} Another reason to take a probabilistic approach is to provide participation incentives~\citep{BBH15b,ABM17a,ALR18a}. For the example described above, at least one of the agents appears to have no strict incentive to participate if the decision is made deterministically. On the other hand, probabilistic rules can be designed that give each voter the ability to at least make an epsilon difference (in expectation) to the outcome. 
	\item \textbf{Achieving strategyproofness without resorting to dictatorship} Some of the most striking results in social choice give the message that if one wants agents to have incentives to report truthfully, then one has to resort to dictatorship. In our running example it would mean selecting the preferred social outcome of one pre-specified agent. However, with a probabilistic or time-sharing approach, one can still achieve strategyproofness and also circumvent the prospect of a single agent with over-riding power~\citep{ABBM14a,CLPP12a,Proc10a,PrTe13a}. 
	\item \textbf{Achieving stability} In much of the social choice and cooperative game theory literature, a theme of results involves the lack of outcome which satisfies an appropriate notion of stability. In voting, the most prominent result within this theme is Condorcet's theorem says that it can be possible that for any given social outcome, a majority of people prefer another outcome. However, Condorcet's cycles vanish when the probabilistic `maximal lottery' rule is used~\citep{ABBH12a,Bran13a}.\footnote{The argument for the existence of such a lottery invokes von Neumann's minimax theorem. } Similarly, core stable outcomes may not exist
for general settings such as hedonic coalition formation in which agents have preferences over coalitions they are members of. However, if we allow for probabilistic outcomes or for time-sharing arrangements, there exist stable outcomes~\citep{AhFl03a}.
	\item \textbf{Circumventing impossibility results} Social choice is at times notorious for some of the bleak impossibility results in its literature. Several results showing that no apportionment method simultaneously satisfies basic monotonicity axioms~\citep{Youn94a}. However, these problems disappear if a bit of randomisation is used. Similarly, there are results pointing out that no deterministic voting rule satisfies some basic consistency properties. However, this is not anymore the case if one uses the \emph{maximal lotteries} randomised voting rule~\citep{Bran13a}.\footnote{For further discussion on probabilistic approaches to circumvent impossibility results in voting, we refer to the survey by \citet{Bran17a}.}
	\item \textbf{Better welfare guarantees}  
	When considering cardinal preferences over outcomes, a probabilistic approach may achieve better approximation  welfare guarantees while simultaneously achieving other axiomatic properties~\citep{ABP15a,AnPo16a,Proc10a}. In some cases, randomization may allow for better welfare or ex ante Pareto improvement while satisfying stability constraints~\citep{Manj16a}.
\end{enumerate}

\section{Research challenges and opportunities}

I outlined several advantages of considering a probabilistic approach. 
%There are several challenges as well. 
At an abstract level, resorting to randomisation means that one can consider the full continuous space of outcomes. This can both be a challenge as well as an opportunity for new and exciting research. 

\begin{enumerate}
	
	\item \textbf{Formalizing and exploring a range of solution concepts and axioms} 
	When considering probabilistic or time sharing approches, there are several ways in which a solution concept for deterministic settings can be extended to probabilistic settings. Take for example pairwise stability for the classic stable matching problem in which we want to pair men and women in a way so that no man and woman not paired with each other want to elope. When considering probabilistic outcomes, there is a hierarchy of stability concepts that are all generalisations of deterministic stability~\citep{AzKl17a,DoYi16a,KeUn15a,Manj16a}. Understanding the nature and structure of these properties is already a  significant research direction. More importantly, the potential to generalize important axioms based on stability, Pareto optimality, and fairness in several different ways gives useful creative leeway for institution designers for exploring the tradeoffs and compatibility between different levels of properties.
	
	 Similar to the potential of defining a several levels of axiomatic properties, one can explore different levels of properties of mechanisms. A case in point is strategyproofness and participation incentives.  
	
	% \footnote{In many of the settings, agents express preferences over discrete outcomes. There are multiple ways to extend preferences over discrete outcomes to preferences over randomized outcomes. Each method of preference extension leads to a corresponding notion of axiomatic property.}
	\item \textbf{Eliciting, representing, and reasoning about preferences}
	In most voting or matching settings, agents express preferences over deterministic outcomes. As we move from deterministic to probabilistic settings, there is an interesting challenge to elicit and represent agents' risk attitudes towards different lotteries. One possible approach that involves compact preferences is  to extend the agents' preferences over discrete outcomes to preferences over probabilistic outcomes by \emph{lottery extension} methods such as first order stochastic dominance~\citep{Bran13b,Bran17a,BoMo01a,ABBH12a,Cho15a}.  %,ChDo16a
	% \item \textbf{Formaling properties of mechanisms} Similar to the potential of defining a several levels of axiomatic properties, one can explore different levels of properties of mechanisms. A case in point are strategyproofness and participation incentives. 
	\item \textbf{Designing time-sharing mechanisms} Since modeling time-sharing is one of the most important motivations of probabilistic social choice, it is important to come up with compelling time-sharing mechanisms. Although voting has been studied for decades, a probabilistic perspective leads to interesting and meaningful new voting rules~\citep[see \eg][]{AzSt14a}. %Aziz13
	When allowing for fractional outcomes, several well-known mechanisms such as Gale-Shapley Deferred Acceptance or Gale's Top Trading Cycles need to be generalized~\citep{KeUn15a,AtSe11a,BoMo04a}. % . It is only recently that researchers have started working on these generalizations
	 
	\item \textbf{Efficiency issues} When trying to achieve fairness via randomization, a straightforward approach is to uniformly randomize over reasonable deterministic outcomes or reasonable deterministic mechanisms. However, such a naive approach such as randomizing over Pareto optimal alternatives can lead to loss of ex ante efficiency.  This phenomenon is starkly highlighted by the fact that random serial dictatorship that involves uniform randomization over the class of serial dictatorships can lead to unambiguous loss of welfare~\citep{BoMo01a,ABBH12a}. This issue motivates the need to design  interesting new mechanisms that are not victim to such a phenomenon. 
	
%	In other cases, randomisation may allow for higher welfare or ex ante Pareto improvement~\citep{Manj16a}. Finding such high welfare outcomes that satisfy other stability requirements can be algorithmically challenging. 
		%\item \textbf{Exploring the full space of outcomes} At an abstract level, resorting to randomisation means that one can consider the full continuous space of outcomes. This can be both a challenge as well as an opportunity. 
		
	\item \textbf{Computational complexity} Generally speaking, optimising in continuous environments is computationally more tractable than in  discrete environments. However, when considering time-sharing outcomes that are implicitly  convex combinations of a potentially exponential number of discrete outcomes, computing the time shares can be a computationally arduous task~\citep{ABBM14a}. Therefore, when formulating time-sharing mechanisms for different social choice settings, computational complexity rears its head as a potential challenge as well an opportunity for innovative algorithmic research. 
	\item \textbf{Instantiating a lottery} As mentioned earlier, uniformly randomizing over desirable outcomes can result in loss of  efficiency or computational intractability. Therefore, mechanisms may first use an alternative way to find an \emph{expected} `fractional' outcome---say a weighted matching signifying probabilities for partnerships. If approach (1) is being used, then the expected outcome needs to be instantiated via a concrete lottery. Finding a suitable lottery is trivial in single-winner voting and easy for simple assignment settings\footnote{Any fractional bipartite matching can be represented as a convex combination of discrete matchings via Birkhoff's algorithm.} but can become a challenge for richer settings with more complex constraints~\citep{BCKM12a}. When instantiating a lottery, an interesting challenge that arises is to instantiate over deterministic outcomes that \emph{also} satisfy some weaker notions of stability, fairness, or other properties~(see e.g., \citep{TST01a,AkNi17a}).   
% \item \textbf{Dealing with weak preferences}  Much of the social choice literature assumes that agents express linear orders. The approach does not capture dichotomous preferences~\citep{BoMo01a,BoMo04a,BMS05a} or certain realistic scenarios. Issues become significantly muddier when dealing with ties. Even simple and canonical mechanisms such as random dictatorships~\citep{Gibb77a} can be extended to the case of weak preferences in several compelling ways depending on the goals.
% For example, under linear orders in voting, the class of strategyproof
\end{enumerate}

To conclude, a probabilistic approach to social choice in particular voting, allocation, matching, and coalition formation leads to several interesting research questions and directions.

\paragraph{Acknowledgements}

The author is supported by a Julius Career Award. He thanks Gabrielle Demange, J{\"{o}}rg Rothe, Nisarg Shah, Paul Stursberg, Etienne Billette De Villemeur, and Bill Zwicker for useful feedback.  He thanks all of his collaborators on this topic  in particular Florian Brandl, Felix Brandt, and Bettina Klaus  for several insightful discussions. 
Finally he thanks Herv{\'{e}} Moulin and Bill Zwicker for encouraging him to write the chapter.

\end{document}